\documentclass[9pt,twocolumn,twoside]{pnas-new}
\templatetype{pnasresearcharticle} % Choose template 
\def\pr{{\rm Pr}}
\def\Red{{\rm  Red}}
\def\Green{{\rm  Green}}
\def\calS{{\mathcal S}}

\def\sd{{\rm SD}}
\def\upsilon{{\Upsilon}}
\def\bi{\begin{itemize}}
\def\ei{\end{itemize}}
\def\Real{\mathbb{R}}
\def\iid{{\ {\buildrel \rm{iid}\over \sim}\ }}
\newcommand{\D}[1]{\mathrm{d}{#1}}
\def\Deq{{\ {\buildrel {\rm D}\over =}\ }}

\title{An Unethical Optimization Principle}

\author[a, 1]{Nicholas Beale}
\author[b]{Heather Battey} 
\author[c]{Anthony C. Davison}
\author[d]{Robert S. MacKay}

\affil[a]{Sciteb Ltd., 23 Berkeley Square, London W1J 6HE, UK ({\tt nicholas.beale@sciteb.com})}
\affil[b]{Department of Mathematics, Imperial College London, 180 Queen's Gate, London SW7 2AZ, UK}
\affil[c]{Institute of Mathematics, Ecole Polytechnique F\'ed\'erale de Lausanne, Station 8, 1015 Lausanne, Switzerland}
\affil[d]{Mathematics Institute, Zeeman Building, University of Warwick, Coventry CV4 7AL, UK}

\leadauthor{Beale} 

\significancestatement{This paper formulates the Unethical Optimization Principle for AI and analytically quantifies the risk amplification involved.  Under mild assumptions we show that an AI is almost certain to adopt an unethical strategy when the returns are Gaussian or have a similar thin-tailed distribution, and that although the probability that such a strategy is adopted decreases as the returns become heavier-tailed, it is still appreciably higher than the incidence of unethical strategies in the strategy space as a whole.  The implications for owners and regulators are that special care must be taken, but the Principle can also be used to help root out ethically problematic strategies
}
\authorcontributions{NB had the initial idea, formulated the Principle, co-wrote the paper, derived equation [4] from the analysis by AD and equation [5]. HB indicated that the extremal types theorem could be used to quantify the risk in wide generality. RM did the initial analysis, leading to formulating the problem in terms of the Odds Ratio. AD provided most of the analysis and co-wrote the paper. All authors contributed importantly to the review and editing of the paper, and did extensive background analysis.}
\authordeclaration{The authors declare no conflict of interest.}
\correspondingauthor{\textsuperscript{1}To whom correspondence should be addressed. E-mail: nicholas.beale@sciteb.com}


\begin{abstract}
If an artificial intelligence aims to maximise risk-adjusted return, then under mild conditions it is disproportionately likely to pick an unethical strategy unless the objective function allows sufficiently for this risk. Even if the proportion $\eta$ of available unethical strategies is small, the probability $p_U$ of picking an unethical strategy can become large; indeed unless returns are fat-tailed $p_U$ tends to unity as the strategy space becomes large. We define an Unethical Odds Ratio Upsilon ($ \upsilon $)  that allows us to calculate $p_U$   from $ \eta $,  and we derive a simple formula for the limit of $\upsilon$ as the strategy space becomes large. We give an algorithm for estimating $ \upsilon $  and $p_U$ in finite cases and discuss how to deal with infinite strategy spaces. We show how this principle can be used to help detect unethical strategies and to estimate $\eta$. Finally we sketch some policy implications of this work.
\end{abstract}

\dates{This manuscript was compiled on 12 November 2019}
\doi{\url{www.pnas.org/cgi/doi/10.1073/pnas.XXXXXXXXXX}}

\begin{document}

\maketitle
\thispagestyle{firststyle}
\ifthenelse{\boolean{shortarticle}}{\ifthenelse{\boolean{singlecolumn}}{\abscontentformatted}{\abscontent}}{}

% If your first paragraph (i.e. with the \dropcap) contains a list environment (quote, quotation, theorem, definition, enumerate, itemize...), the line after the list may have some extra indentation. If this is the case, add \parshape=0 to the end of the list environment.
\dropcap{A}rtificial intelligence (AI) is increasingly deployed in commercial situations. Consider for example using AI to set prices of insurance products to be sold to a particular customer. There are legitimate reasons for setting different prices for different people, but it may also be profitable to ``game''  their psychology or willingness to shop around. The AI has a vast number of potential strategies to choose from, but some are unethical --- by which we mean, from an economic point of view, that there is a risk that stakeholders will apply some penalty, such as fines or boycotts, if they subsequently understand that such a strategy has been used.  Such penalties can be huge:~although these happened too early for an AI to be involved, the penalties levied on Banks for misconduct are  currently estimated to be over \$276 billion (see SI). In an environment in which decisions are increasingly made without human intervention, there is therefore a strong incentive to know under what circumstances AI systems might adopt unethical strategies. Society and governments are closely engaged in such issues. Principles for ethical use of AI have been adopted at national \citep{HMG:2019} and international \citep{OECD:2019} levels and the whole area of AI Ethics is one of very considerable activity \citep{Bostrom.Yudkowsky:2011,Dignum:2018}.
%\footnote{Nicholas: The earliest cited `paper' on the subject is Asimov (1942) according to \citep{Bostrom.Yudkowsky:2011}. \parshape=0} 

Ideally there would be no unethical strategies in the AI's strategy space. But the best that can be achieved may be to have only a small fraction $\eta$ of such strategies being unethical. Unfortunately this runs up against the Unethical Optimization Principle, which we formulate as follows.

%If an AI aims to maximize Risk-Adjusted Return, then under mild conditions it is disproportionately likely to pick an unethical strategy unless the Objective Function allows sufficiently for this risk.

\textbf{If an AI aims to maximise risk-adjusted return, then under mild conditions it is disproportionately likely to pick an unethical strategy unless the objective function allows sufficiently for this risk.}

\section*{Problem formulation}

% ANTHONY'S PROPOSED VERSION =========================

The following is a deliberately oversimplified representation that emphasises certain aspects and ignores others.  Consider an AI that is searching a strategy space $\calS$ for a strategy $s$ that maximises the risk-adjusted return for its owners.  It does this by attempting to maximise its estimate of apparent risk-adjusted return function $A(s)$,  which we treat as random because it is based on potentially noisy data --- for example data from existing clients who are themselves taken from a much larger number of potential clients.  However, unknown to the AI, certain strategies in $\calS$ would be considered unethical by stakeholders, who in the future may impose a penalty for adopting them. Such penalties may be fines, reparations/compensation or boycotts: what they have in common from our point of view is that they have a non-zero risk-adjusted expected cost which we denote by $C(s)$. We will call the subset of  $\calS$ for which $C(s) >0$ ``unethical''  or $\Red$, and the complementary subset, for which $C(s) = 0$, ``ethical''  or $\Green$.   Hence the true risk-adjusted return $T(s)$ may be expressed as
\begin{equation}\label{eq0}
T(s)= A(s) - C(s) + Q(s), 
\end{equation}
where the `error' $Q(s)$ accounts for other differences between $T(s)$ and $A(s)$ even when $C(s)=0$, due to imperfections in the algorithm’s ability to predict the future accurately.
%I suggest we remove these footnotes \footnote{The extent to which T, A, C and Q are Random Variables raises some subtle points. Most AI systems are trained on large amounts of data but apply algorithms which, given the training at a point in time, are deterministic. Therefore for a given value of s, A(s) is in principle deterministic at any point in time. By contrast T, C and hence Q depend on unknowable future states of the world and are hence 'genuinely random'. If we write $E_f[X]$ to mean the expectation over future states of the world then we could re-write [1] as $A(s) = E_f[T(s)] + E_f[C(s)] - E_f[Q(s)]$.}\footnote{Comment from AD on this footnote: I think we should keep this sort of discussion to a minimum.  Apart from anything else, the algorithms can be random (stochastic gradient descent is very common), so the optimum found may be random.  We are proposing a model for the optimisation, and provided the model is plausible/reasonable, I don't think we need to justify things in too much detail.  But we do need the notation to be coherent and not too heavy. For this reason I would prefer to avoid extra subscripts.  Also, I don't know of any book/paper in which random variables are `Random Variables'.  We're halfway to Trumpsville \ldots}

Let $p_U= \pr(s^\ast\in \Red)$ denote the probability that the chosen strategy $s^\ast$ is unethical, and assume there is some  measure on $\calS$ so that one could in principle compute  the proportion $\eta$ of $\calS$ that is Red.  The $\Green$ strategies comprise the remaining proportion $1-\eta$ of $\calS$. Then we  can define an Unethical Odds Ratio, Upsilon, as: 
\begin{equation}
\label{up.eq}
\Upsilon := \dfrac{p_U}{1-p_U}\div \dfrac{\eta}{1-\eta},
\end{equation}
which represents the increase in odds of choosing an unethical strategy by using the AI, relative to choosing a strategy at random.  If $\eta$ is small, then $\Upsilon \approx 1$ will not represent a significant increase in risk due to use of the AI, whereas if $ \Upsilon \gg 1$ then the AI acts as a significant unethical amplifier. If regulation reduces $\eta$ to 0.05  (or 0.01), for example, having $\Upsilon = 10$ would mean $p_U\approx 0.35$  (or 0.09).

Unless there is a difference in the distribution of  $T+Q$ between the Red and Green regions or the mean returns are infinite, the expected risk-adjusted returns in $\Red$ and $\Green$ satisfy 
\begin{align*}
E_s[A\mid \Red] &= E_s[T+Q\mid \Red] + E_s[C\mid \Red] \\
&>E_s[T+Q] =  E_s[A\mid \Green],
\end{align*}
where $E_s[X \mid \Red]$ means the average value of $X$ for $s\in\Red$.  
%\footnote{This will be exact if $E_s[T+Q \mid \Red] = E_s[T+Q]$. If we regard $T+Q$ as random this equality will not be exact because the $\Red$ and $\Green$ subsets will be different samples from the same underlying population. However under very mild conditions, as $S$ becomes large $E_s[T+Q \mid \Red] - E_s[T+Q \mid \Green]$ becomes very small, so this inequality will hold if $E_s[C \mid \Red]$ is not very small. See SI for further discussion.} 

Moreover if $C(s)$ varies within $\Red$,  
%if $C$ is uncorrelated with $T+Q$, then 
the corresponding standard deviations will satisfy $\sd(A\mid \Red) > \sd(A\mid \Green)$. Thus returns for strategies in $\Red$ will have higher means and variabilities than those in $\Green$.  Suppose that the mean estimated return in $\Red$ is $\Delta>0$ larger than that in $\Green$, and that the estimated standard deviation in $\Red$ is a factor $1+\gamma$ larger than that in $\Green$.   The trade-off between returns from ethical and unethical strategies will depend on $\eta$, $\Delta$ and $\gamma$ and on the tail of the  distribution of returns. 

\subsection*{Asymptotic strategy space} Suppose that the strategy space $\calS$ contains $S$ strategies,  of which $m=\eta S$ are unethical and $n=(1-\eta)S$ are ethical. Let $M_R$ and $M_G$ respectively denote the maximum returns for strategies in  $\Red$ and $\Green$.  In many cases the maximum $M_n$ of a random sample of size $n$ from a distribution  $F$ can be renormalized using sequences $\{a_n\}>0$ and $\{b_n\}\subset\Real$ in order that $(M_n-b_n)/a_n$ converges as $n\to\infty$ to a limiting random variable $X$ having a generalized extreme-value distribution.  This distribution has a tail index parameter $\xi$ that controls the weight of its right-hand tail, with increasing $\xi$ corresponding to fatter tails; it includes the Gumbel distribution $\exp\{-\exp(-x)\}$ as a special case for $\xi=0$. Following the discussion above, we can write $M_R=\Delta+(1+\gamma)M_m$ and $M_G=M_n$, where $M_m$ and $M_n$ are respectively the maxima of $m$ and $n$ mutually independent variables from $F$, and we suppose that $(M_m-b_m)/a_m$ and $(M_n-b_n)/a_n$ converge to variables $X$ and $Y$, which are independent and have the same generalized extreme-value distribution. In the Supporting Information we obtain general expressions for the limiting probability $p_U$ under mild conditions, and compute $p_U$ and $\upsilon$ for some special cases:
\bi
\item if $F$ is Gaussian, then the limiting variables $X$ and $Y$ are Gumbel, and $\Upsilon \to \infty$ if $\Delta$, $\gamma$ or both are positive; 

%
%
%The fact that $a_n\to 0$ means that the distribution of the maximum becomes more and more concentrated around $b_n$ as $n$ increases, which might be interpreted as meaning that as the number of independent actors in a market increases, the best performers among them become increasingly indistinguishable.  Whether the actors in an investment market are independent seems moot, but the same limit emerges provided the $Z_j$ are not too dependent \citep{Leadbetter.Lindgren.Rootzen:1983}.  Hence if  $\mu>0$ then $\pr( Y_m>X_n) \to 1$ for any choice of $\gamma$: if some of the actors have a clear advantage, then one of them is guaranteed to perform best overall.  If on the other hand $\mu=0$, and there is no systematic advantage, then the limiting probability $\pr\{ 2\log \eta + (1+\gamma)Y > X\} $ will be positive for any $p$ and $\gamma$ and is decreasing as $p$ decreases for fixed $\gamma$.   In this case $Y$ could beat $X$, but is not certain to do so.  It would not be hard to estimate the probabilities for a range of values of $p$ and $\gamma$;  I think this can be done explicitly when $\gamma=0$, but would require some (fairly easy) numerical work otherwise.  

\item if $F$ is lognormal or exponential, then the limiting variables $X$ and $Y$ are Gumbel and $\upsilon \to\infty$ if $\gamma>0$;

\item if $F$ is Pareto, i.e., $F(x) = 1-x^{-\nu}$ for $x>1$ and $\nu>0$, then $X$ and $Y$ have Fr\'echet distributions with tail indexes $\xi=1/\nu$, and 
\begin{equation}
\label{pU.eq}
\lim_{S\to\infty} p_U = \dfrac{\eta (1+\gamma)^\nu}{1 -\eta + \eta (1+\gamma)^\nu },
\end{equation}
which yields 
\begin{equation}
\label{eq1}
\upsilon \to \upsilon^* =(1+\gamma)^\nu \quad \mbox{ as } \quad S\to\infty;
\end{equation}
and 
\item if $F$ is Student $t$ with $\nu$ degrees of freedom, then the Pareto limit applies. 
\ei

\begin{figure}%[tbhp]
\centering
\includegraphics[width=1.00\linewidth]{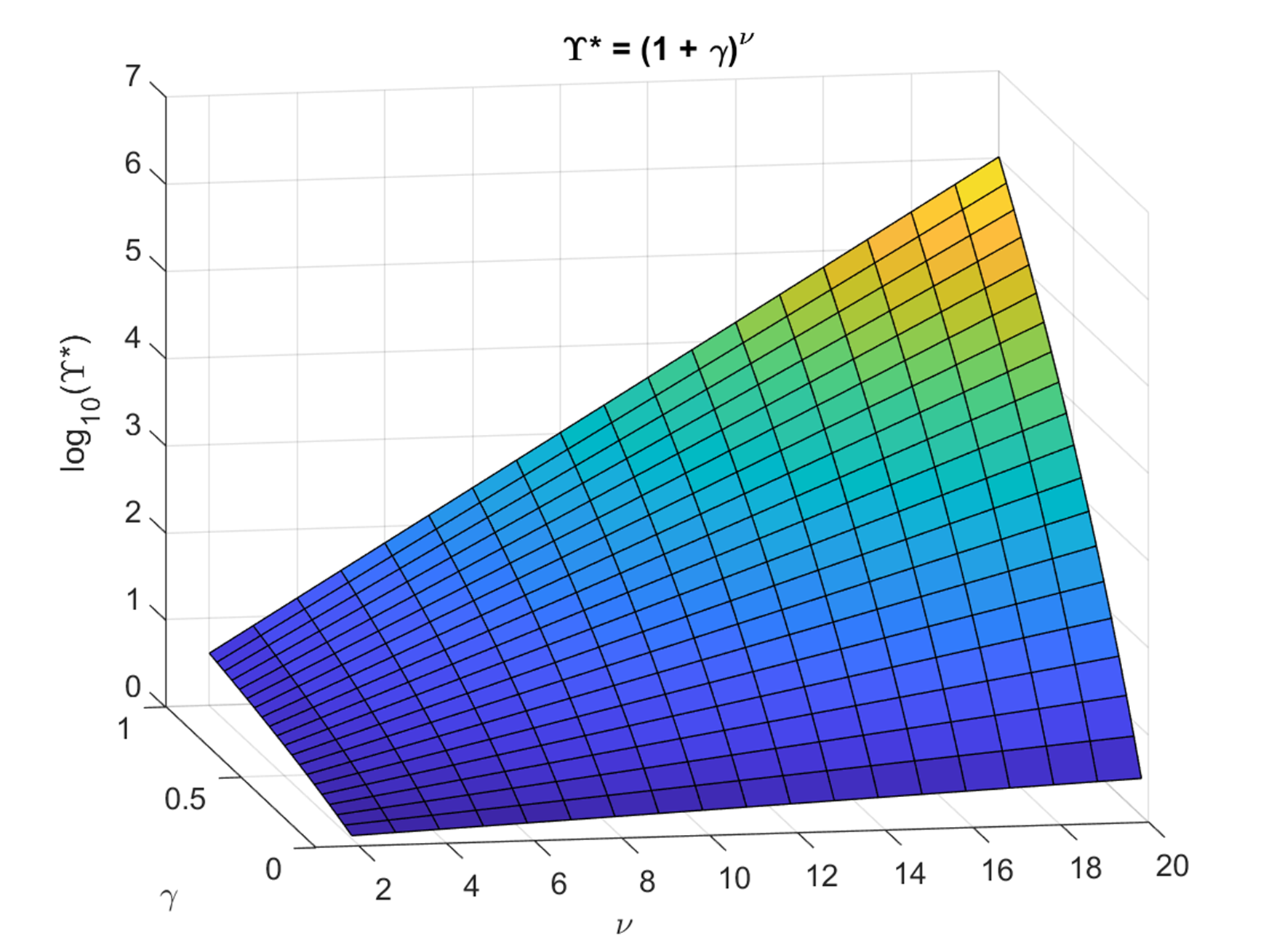}
\caption{Dependence of the asymptotic unethical odds ratio $\upsilon^*$ on tail index $\nu$ and additional volatility $\gamma$.}
\label{fig2}
\end{figure}

The significance of these results is that if the strategy space is large, then unless the distribution of the returns is fat-tailed, as in the cases of the Pareto or $t$ distributions, a responsible regulator or owner should be extremely cautious about allowing AI systems to operate unsupervised in situations with real consequences. If the returns are fat-tailed, then~\eqref{eq1} gives some idea of the risk of using an unethical strategy.  

Figure~\ref{fig2} shows how the tail index $\nu$ influences~\eqref{eq1} in the heavy-tailed case.  If $\nu=7$, for example, then $\upsilon^* \approx 1.4$ for $\gamma=0.05$ and  $\upsilon^*\approx 17$ for $\gamma=0.5$.  For large $\gamma$ the value of $\upsilon^*$ rises rapidly with $\nu$, and it remains small for all $\nu$ only when $\gamma\approx 0$. 

\subsection*{Results for finite strategy space} 

For large but finite $S$ there is a simple and widely-applicable algorithm to estimate $\upsilon$. Numerical experiments show that its limiting value $\upsilon^*$ is reached quite rapidly for fat-tailed distributions, whereas $\upsilon$ grows roughly as $\log S$ for Gaussian returns. 

\begin{SCfigure*}[\sidecaptionrelwidth][t]
\centering
\includegraphics[width=1.7\linewidth]{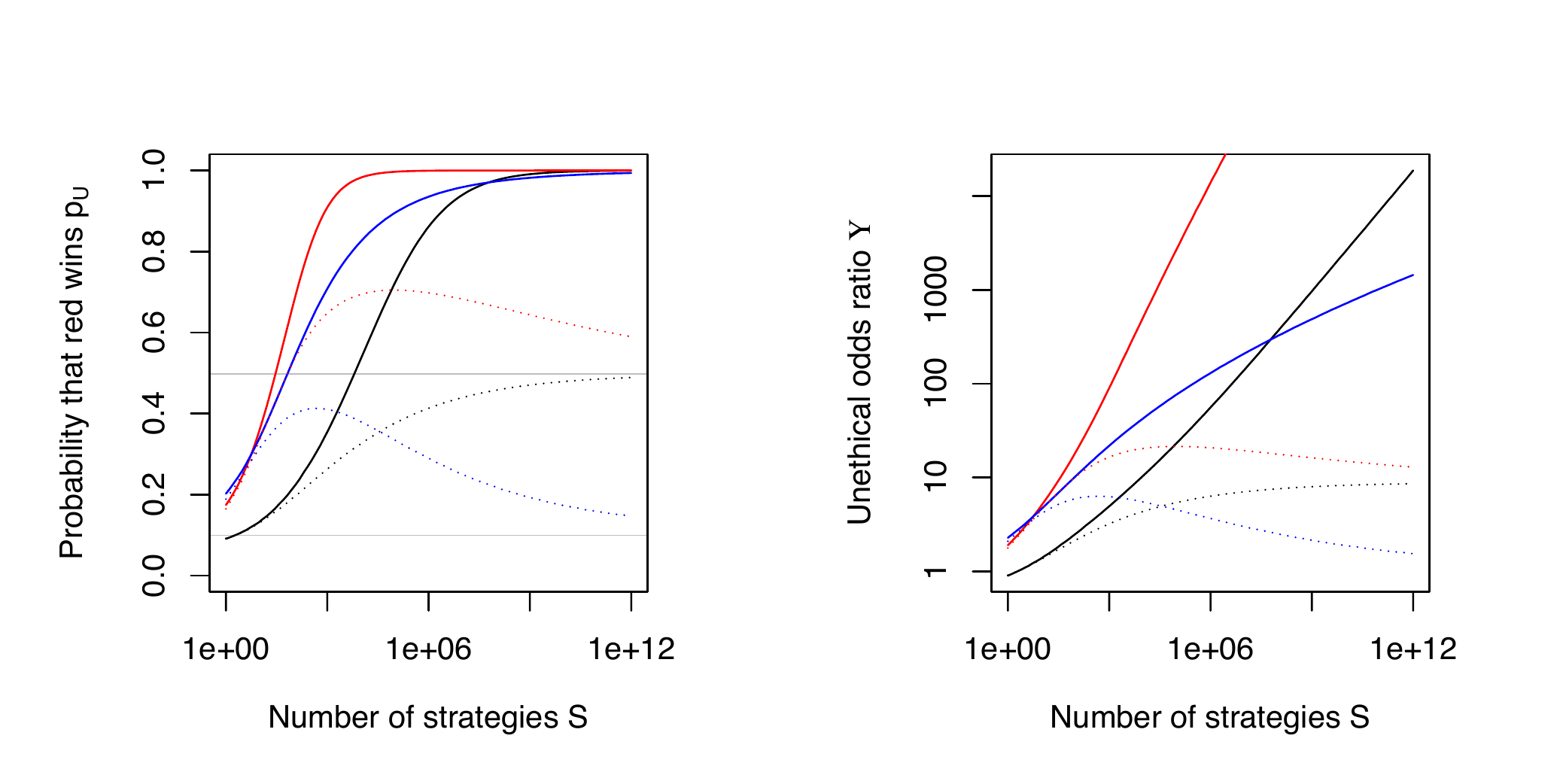}
\caption{Dependence of probability $p_U$ and Unethical Odds Ratio $\upsilon$ on size of strategy space $\calS$ for normal distribution (solid) and $t_{12}$ distribution (dots) when $\eta=0.1$: $\gamma=0.2$, $\Delta=0$  (black); $\gamma=0.2$, $\Delta=0.5$ (red); $\gamma=0$,  $\Delta=0.5$ (blue). The grey horizontal lines in the left-hand panel show the limiting probabilities from~\eqref{pU.eq}.}
\label{fig1}
\end{SCfigure*}

% \begin{figure}%[tbhp]
%\centering
%\includegraphics[width=1.00\linewidth]{fig2.pdf}
%\caption{Dependence of probability $p_U$ and Unethical Odds Ratio $\upsilon$ on size of strategy space $\calS$ for normal distribution (solid) and $t_{12}$ distribution (dots) when $\eta=0.1$: $\gamma=0.2$, $\delta=0$  (black); $\gamma=0.2$, $\delta=0.5$ (red); $\gamma=0$,  $\delta=0.5$ (blue). The grey horizontal lines in the left-hand panel show the limiting probabilities from~\eqref{pU.eq}.}
%\label{fig1}
%\end{figure}

Figure~\ref{fig1} shows how the finite-sample unethical odds ratio $\upsilon$ depends on $S$ for some special cases.  In the Gaussian case the probabilities approach unity most rapidly when the volatility is inflated, i.e., $\gamma>0$, and the Unethical Odds Ratio appears to be ultimately log-linear in $\log S$.  In the case of Student $t$ returns with $\nu=12$ degrees of freedom, the probabilities overshoot their asymptotic values when $\Delta>0$, and the asymptote~\eqref{eq1} is approached rather slowly.

\subsection*{Infinite strategy spaces and correlated returns} 

So far we have discussed finite strategy spaces in which the returns for each strategy are independent. For many purposes this may be enough: if the asymptotic values of $\upsilon$ and $p_U$ are known it may be irrelevant whether $S$ is $10^6$ or $10^{26}$.  However there may be an effective upper bound on $S$ even when $\calS$ is infinite, if $A(s)$ is viewed as a stochastic process with state space $\calS$.  For example, if there is a metric on $\calS$ and there are correlations between neighbouring points. Understanding the best approach in particular cases will depend on knowing the structure of $\calS$ and of $A(s)$, but these are the one part of the system that are essentially under the control of the AI and therefore the least imponderable. The more that is known about $\calS$ and $A(s)$ the better one can estimate the effective value of $S$. We discuss this further in the SI.
% remove this at least for now: One way to tackle  this is to observe how $\max_{s\in\calD} A(s)$ increases for increasingly large samples $\calD$ drawn from $\calS$. Once the sample size $D=|\calD|$ greatly exceeds the effective number of independent strategies in $\calS$, the rate at which $\max_{s\in\calD}  A(s)$ grows will tail off faster than would be expected from the observed distribution of $A(s)$.\footnote{how do we know this distribution? NB from sampling.} This puts a finite upper bound on the effective value of $S$ in infinite strategy spaces. 

\subsection*{Estimating the parameters} 

The Unethical Optimisation Principle can help risk managers and regulators to detect unethical strategies. Consider a reasonably large sample $L \subset \calS$. Manually examining $L$ for potential unethical elements may be prohibitively expensive if this requires human judgement. Suppose however that we rank the elements of $L$ by their values of $A(s)$ and focus our attention on the subset $L_k$ with the $k$ largest values of $A(s)$, where $k \ll |L|$. We assume that careful manual inspection can divide this set into $\Red$ and $\Green$ elements and write $\hat p_{U_k} = |L \cap \Red|/k$. By~\eqref{up.eq} we then have an estimator 
\begin{equation}
\label{eta.eq}
\hat \eta_k = \dfrac{\hat p_{U_k}}{(1-\hat p_{U_k})\Upsilon + \hat p_{U_k}},
\end{equation}
which allows a rough estimate of $\eta$ given $\Upsilon$ and $\hat p_{U_k}$. Perhaps more importantly, focusing on $L_k$ to find examples of unethical strategies that might be adopted not only weeds out those most likely to be used, but will help develop intuition on where problems might be found. Observing the bulk distribution of $A(s \mid s \in L)$ gives an idea of overall shape of $A(s)$ and an idea of $\nu$.  To generate reasonably robust estimates of $\gamma$ and $\Delta$ it will generally be necessary to do some more manual inspection of another subset of $L$ to determine $\Red$ and $\Green$ elements but this can be relatively small if well targeted. Details are discussed in the SI.

\subsection*{Implications}  Practical advice to the regulators and owners of AI is to  sample the strategy space and observe whether the returns $A(s)$ have a fat-tailed distribution. If not, then the ``optimal'' strategies are likely to be unethical regardless of the value of $\eta$.  If, however, the observed return distribution is fat-tailed, then  the tail index $\nu$ can be estimated using standard techniques \citep{Embrechts.Kluppelberg.Mikosch:1997,Coles:2001} and $\eta$ can be estimated as discussed above. However, it would be unwise to place much faith in the precision of such estimates:~there are so many imponderable factors that the main point is to avoid sailing close to the wind. In addition the Principle can be used to help regulators, compliance staff and others to find problematic strategies that might be hidden in a large strategy space. 

The Principle also suggests that it may be necessary to re-think the way AI operates in very large strategy spaces, so that unethical outcomes are explicitly rejected in the optimisation/learning process. %Research on how this can be achieved is under way, though it may require new concepts to extend conventional economics. We look forward to participating in, and encouraging, such work.

%\section*{Closing} This note introduces the Unethical Optimisation Principle and provides a simple formula to estimate its impact, as well as providing code for more detailed exploration. We hope that this quantitative connection between economics,  financial regulation and AI ethics will provide a fruitful basis for discussion and for further research.

\acknow{This work was supported by the Swiss National Science Foundation,  the UK Engineering and Physical Sciences Research Council and  Capital International.  We thank Prof He Ping, Deputy Governor Pan Gonsheng and Alex Brazier for organising seminars at Tsinghua School of Economics and Management, the PBOC/SAFE and the Bank of England in March and April 2019 where NB presented the initial ideas that led to this paper. RM is grateful to the Alan Turing Institute for a Fellowship TU/B/000101 that helped enable this collaboration. We thank Andrew Bailey, Karen Croxson, and Wolfram Peters for helpful discussions.}

\showacknow{} % Display the acknowledgments section

% Bibliography
\bibliography{pnas-sample}

\begin{thebibliography}{10}

\bibitem{HMG:2019}
{UK Government Data Ethics Framework} (2019).

\bibitem{OECD:2019}
{OECD Principles on AI} (2019).

\bibitem{Bostrom.Yudkowsky:2011}
N Bostrom, E Yudkowsky, Ai ethics in {\em {Cambridge Handbook of Artificial
  Intelligence}}, eds.{} W Ramsey, K Frankish.
\newblock (Cambridge University Press), (2011).

\bibitem{Dignum:2018}
V Dignum, {Ethics in artificial intelligence: Introduction to the special
  issue}.
\newblock {\em\protect\JournalTitle{Ethics and Information Technology}}
  \textbf{20}, 1--3 (2019).

\bibitem{Embrechts.Kluppelberg.Mikosch:1997}
P Embrechts, C Kl{\"u}ppelberg, T Mikosch, {\em {Modelling Extremal Events for
  Insurance and Finance}}.
\newblock (Springer, Berlin), (1997).

\bibitem{Coles:2001}
SG Coles, {\em {An Introduction to Statistical Modeling of Extreme Values}}.
\newblock (Springer, New York), (2001).

\bibitem{Megaw:2019}
N Megaw, {The UK’s slow-burn £50bn banking scandal}.
\newblock {\em\protect\JournalTitle{Financial Times}} (2019).

\bibitem{Leadbetter.Lindgren.Rootzen:1983}
MR Leadbetter, G Lindgren, H Rootz{\'e}n, {\em {Extremes and Related Properties
  of Random Sequences and Processes}}.
\newblock (Springer, New York), (1983).

\bibitem{Fisher.Tippett:1928}
RA Fisher, LHC Tippett, {Limiting forms of the frequency distributions of the
  largest or smallest member of a sample}.
\newblock {\em\protect\JournalTitle{Proceedings of the Cambridge Philosophical
  Society}} \textbf{24}, 180--190 (1928).

\bibitem{Davison.Smith:1990}
AC Davison, RL Smith, {Models for exceedances over high thresholds (with
  Discussion)}.
\newblock {\em\protect\JournalTitle{Journal of the Royal Statistical Society,
  series B}} \textbf{52}, 393--442 (1990).

\bibitem{Davison:2003}
AC Davison, {\em {Statistical Models}}.
\newblock (Cambridge University Press, Cambridge), (2003).

\bibitem{R2018}
{R Core Team}, {\em R: A Language and Environment for Statistical Computing} (R
  Foundation for Statistical Computing, Vienna, Austria), (2018).

\bibitem{Coles.Heffernan.Tawn:1999}
SG Coles, J Heffernan, JA Tawn, Dependence measures for extreme value analyses.
\newblock {\em\protect\JournalTitle{Extremes}} \textbf{2}, 339--365 (1999).

\bibitem{Davis.Mikosch:2009:Ext}
RA Davis, T Mikosch, The extremogram: A correlogram for extreme events.
\newblock {\em\protect\JournalTitle{Bernoulli}} \textbf{15}, 977--1009 (2009).

\bibitem{Leadbetter:1991}
MR Leadbetter, {On a basis for `Peaks over Threshold' modeling}.
\newblock {\em\protect\JournalTitle{Statistics \& Probability Letters}}
  \textbf{12}, 357--362 (1991).

\end{thebibliography}
\pagebreak
\section*{Supporting Information Appendix (SI)}

\subsection*{Recent penalties in Financial Services}
The Financial Times listed \citep{Megaw:2019} the major sets of fines and penalties levied on Western Banks for various forms of misconduct. There were 11 types of misconduct and the fines and penalties totaled \$276Bn. Penalties (including compensation) for Payment Protection Insurance totaled \$62Bn and was the second largest category.

\subsection*{Derivation of limiting $p_U$}

The extremal types theorem \citep[Theorem~1.4.2]{Leadbetter.Lindgren.Rootzen:1983} implies that in wide generality, the maximum $M_n$ of a random sample $Z_1,\ldots, Z_n$ with cumulative distribution function $F$  may be renormalized using sequences $\{a_n\}>0$ and $\{b_n\}\subset\Real$ in order that $(M_n-b_n)/a_n$ converges as $n\to\infty$ to a limiting random variable $X$ having a generalized extreme-value distribution.  A simple sufficient condition for this is that $F(x)$ is twice continuously differentiable with density $f(x)$ and that the reciprocal hazard function $r(x) = \{1-F(x)\}/f(x)$ is such that $r'(x)$ converges to a constant $\xi$ as $x$ approaches the upper support point $x^*$ of $f$.  Then we can take $b_n=F^{-1}(1-1/n)$, $a_n=r(b_n)>0$ and the distribution of $X$ is 
\begin{equation}\label{GEV.eq}
G_\xi(x) = \exp\left\{ - (1+\xi x)^{-1/\xi}_+\right\}, \quad x\in\Real,
\end{equation}
where $a_+=\max(a,0)$; setting $\xi=0$ gives the Gumbel distribution $G_0(x) = \exp\{-\exp(-x)\}$. The quantity $\xi$, sometimes called the tail index, typically satisfies $|\xi|<1/2$, with smaller values corresponding to lighter tails.  If $\xi<0$, then the limiting density has an upper support point at $-1/\xi$, whereas if $\xi\geq 0$ then the limiting density has no finite upper support point, so the limiting random variable has no upper bound. 

This implies that we can write $M_n \approx b_n + a_n X$ for sufficiently large $n$, where the quality of the approximation  depends on $F$; it has long been known that the convergence is extremely slow for Gaussian variables  \citep{Fisher.Tippett:1928}.  A result of Khintchine \citep[Theorem~1.2.3]{Leadbetter.Lindgren.Rootzen:1983} implies that if $m=\eta S$ and $n=(1-\eta)S$ for some fixed $\eta\in(0,1)$, then as $S\to\infty$,
\begin{eqnarray*}
%\label{eq11}
\dfrac{b_m-b_n}{a_n} &\to&\beta_\eta =\dfrac{\{\eta/(1-\eta)\}^\xi -1}{\xi}, \\
\dfrac{a_m}{a_n} &\to& \alpha_\eta= \left(\dfrac{\eta}{1-\eta}\right)^\xi, 
\end{eqnarray*}
with $\beta_\eta = \log\{\eta/(1-\eta)\}$ when $\xi=0$. 

To apply these results, let $M_G$ denote the maximum of independent random variables $Z_1,\ldots, Z_n$ with common distribution function $F$, which represent the returns of ethical, Green, strategies, and  suppose that $(M_G-b_n)/a_n$ converges in distribution to a random variable $X$ as $n\to\infty$.  Let $M_R$ denote the maximum of $m$ independent random variables $\Delta+(1+\gamma)Z'_j$ representing the returns of unethical, red,  strategies.  We suppose that $Z_1',\ldots, Z_m'$ is a random sample from $F$ and that  $\Delta\geq 0$ and $\gamma\geq 0$ quantify the increase in mean return and in volatility for unethical returns. We briefly discuss the case where the $Z_j$ and $Z_j'$ have different distributions below.  Then 
$$
M_R\Deq \Delta + (1+\gamma) \max(Z'_1,\ldots, Z'_m),
$$
where $\Deq$ means `has the same distribution as', and as $S\to\infty$,   $\{(M_R-\Delta)/(1+\gamma) - b_m\}/a_m$ will converge in distribution to a random variable $Y$ with  the same distribution as $X$.  

If $m$ is large enough, then we can write $M_R \approx \Delta + (1+\gamma)b_m + a_m(1+\gamma)Y$, and so the probability that the best return from an unethical strategy exceeds the best return from a ethical one satisfies
%$$
%\pr(Y_m > X_n) \approx \pr\left\{ \Delta + (1+\gamma)b_m + a_m(1+\gamma)Y > b_n + a_nX\right\}.
%$$
%When the right-hand side of this expression is rearranged, the limits in \eqref{eq1} imply that 
%<!-- $$ -->
%<!-- \pr\left\{ (b_m-b_n)/a_n +(\Delta+\gamma b_m)/a_n + (a_m/a_n)(1+\gamma)Y >X\right\}. -->
%<!-- $$ -->
%<!-- 
%The limits in \eqref{eq1} imply that  -->
$$
\pr( M_R>M_G) \to \pr\left\{ \beta_\eta +A(\Delta,\gamma,\eta ) + (1+\gamma)\alpha_\eta Y >X\right\}, 
$$
as $S\to\infty$, where $A(\Delta,\gamma,\eta) = \lim_{S\to\infty} (\Delta+\gamma b_m)/a_n$ depends on $\eta$, $\Delta$, $\gamma$ and the normalising sequence for $F$. 
%<!-- If $\eta =1/2$, $\Delta=0$ and $\gamma=0$, then $\alpha_\eta =1$ and $\beta_\eta =0$, so $X$ and $Y$ have the same limiting distribution and the limiting probability is that of $Y>X$, which is 1/2, as expected.   -->

We now discuss the behaviour for large $S$ of 
\begin{equation}
\label{A.eq}
\dfrac{\Delta}{a_n} + \gamma\dfrac{b_m}{a_n} 
= \dfrac{\Delta}{a_n} + \gamma \dfrac{b_m}{a_m}\dfrac{a_m}{a_n}.
\end{equation}
%in the setting above.  
\bi
\item If $x^*<\infty$, then $a_n\to 0$ and $b_m/a_m\to\infty$, so  $A(\Delta,\gamma,\eta )=\infty$.  In this case the distributions of $M_G$ and $M_R$ become more and more concentrated for large $S$, and any advantage for red leads to it beating green with probability one, in the limit, because red returns have a higher upper limit than green ones. 

\item If $x^*=\infty$, then $a_n/b_n=r(b_n)/b_n\to \xi$ as $n\to\infty$, so $b_m/a_n = b_m/a_m\times a_m/a_n\to \xi^{-1}\alpha_\eta$, which is infinite if $\xi=0$.  The behaviour of $\Delta/a_n$ depends on the limit of $a_n=r(b_n)$ as $b_n\to\infty$.  For example, if $F$ is exponential, then $a_n$ converges to a constant, whereas if $F$ is Gaussian, then $a_n\to 0$.  For exponential maxima, therefore, $A(\Delta,\gamma,\eta )$ is infinite if $\gamma>0$, but is finite if $\gamma=0$, for any $\Delta$.  For Gaussian maxima, $\xi=0$ and $a_n\to 0$, so $A(\Delta,\gamma,\eta )=\infty$ if either of $\Delta$ or $\gamma$ is positive, i.e., if there is any systematic advantage for red strategies.
\ei
Other limits  might appear when $\Delta$ and $\gamma$ depend on $S$, but one would need to consider whether this is realistic; for example, this might apply if $\eta\to0$, i.e., red strategies are a vanishingly small fraction of all possible ones.  This does not seem very realistic, since presumably any ethical strategy could be tweaked slightly to make it more profitable but unethical. 

Here are the details for the  special cases in the main text.
\bi
\item If $F$ is Gaussian, then we can take $b_n = (2\log n)^{1/2}$ and $a_n=1/b_n \to 0$, giving $\xi=0$, so $\beta_\eta = \log\{ \eta/(1-\eta)\}$ and $\alpha_\eta =1$.  The limiting variables $X$ and $Y$ are Gumbel, and red will beat green  if either  $\Delta$ or $\gamma$ is positive. 

%The fact that $a_n\to 0$ means that the distribution of the maximum becomes more and more concentrated around $b_n$ as $n$ increases, which might be interpreted as meaning that as the number of independent actors in a market increases, the best performers among them become increasingly indistinguishable.  Whether the actors in an investment market are independent seems moot, but the same limit emerges provided the $Z_j$ are not too dependent \citep{Leadbetter.Lindgren.Rootzen:1983}.  Hence if  $\Delta>0$ then $\pr( Y_m>X_n) \to 1$ for any choice of $\gamma$: if some of the actors have a clear advantage, then one of them is guaranteed to perform best overall.  If on the other hand $\Delta=0$, and there is no systematic advantage, then the limiting probability $\pr\{ 2\log \eta + (1+\gamma)Y > X\}$ will be positive for any $p$ and $\gamma$ and is decreasing as $p$ decreases for fixed $\gamma$.   In this case $Y$ could beat $X$, but is not certain to do so.  It would not be hard to estimate the probabilities for a range of values of $p$ and $\gamma$;  I think this can be done explicitly when $\gamma=0$, but would require some (fairly easy) numerical work otherwise.  

\item If $F$ is log-Gaussian, then we can take $b_n=\exp\{(2\log n)^{1/2}\}$ and $a_n = b_n/(2\log n)^{1/2}$, so $\xi=0$, $\beta_\eta =\log\{\eta/(1-\eta)\}$ and  $\alpha_\eta =1$.  The limiting variables $X$ and $Y$ are Gumbel. Here $a_n\to\infty$ and $b_m/a_n\to\infty$, so red always beats green, owing to its higher volatility.  

\item If $F$ is exponential, then $b_n=\log n$, $a_n=1$ and $\xi=0$, so  $X$ and $Y$ are Gumbel, $\beta_\eta = \log\{\eta/(1-\eta)\}$, $\alpha_\eta =1$ and 
$$
(\Delta+\gamma b_m)/a_n = \Delta + \gamma\log S + \gamma\log \eta 
$$
tends to infinity unless $\gamma=0$: red beats green in the limit owing to its higher volatility.  
%<!-- %If $\gamma=0$ then the limiting probability that red wins is $p/(p+e^{-\Delta})$. In this case one would expect the limiting result to provide reasonable answer for finite samples, because convergence of the maxima to their limits is quite quick. %, because $X-Y$ has a logistic distribution.   -->

\item If $F$ is Pareto, then $b_n=n^{1/\nu}$, $a_n=b_n/\nu$ and $\xi=1/\nu$, so $\beta_\eta =\nu[ \{\eta/(1-\eta)\}^{1/\nu}-1]$, $\alpha_\eta =\{\eta/(1-\eta)\}^{1/\nu}$ and $A(\Delta,\gamma,\eta ) = (1+\gamma)\nu\alpha_\eta$.  Here $X$ and $Y$ have Fr\'echet distributions, $\exp\{-(1+x/\nu)^{-\nu}\}$ for $x>-\nu$, and as  $S\to\infty$, we obtain
\begin{eqnarray}
\pr(M_R > M_G) &\to&% \pr\left\{(1+\gamma)\left(\dfrac{\eta}{1-\eta}\right)^{1/\nu} Y > X\right\} \nonumber \\
%&= &
\dfrac{\eta (1+\gamma)^\nu}{1-\eta  + \eta (1+\gamma)^\nu }.\label{eq2}
\end{eqnarray}
%which equals $\eta$ if $\gamma=0$. 
% ; note that if $X$ and $Y$ are independent Fr\'echet variables with parameter $\nu$, then conditioning on the value of $Y$ yields 
%\begin{eqnarray*}
%\pr( X< \beta Y) &=& \int_0^\infty \nu y^{-\nu-1}\exp\left\{ -y^{-\nu} -(\beta y)^{-\nu} \right\}\, \D{y}\\
%&= & \beta^\nu/(1+\beta^\nu).
%\end{eqnarray*}
Hence $\pr(M_R > M_G) > \eta$ for large $S$ if and only if $\gamma>0$.  This calculation also applies to other distributions with Pareto-like tails, such as the Student~$t$.  Inserting~\eqref{eq2} into~\eqref{up.eq} yields~\eqref{eq1}. 
\ei

The discussion above presupposes that the red and green returns only differ by a location and/or scale shift.  If the limiting variables have the same support but different tail indexes, then the variable with the higher $\xi$ asymptotically dominates the other: if $Y$ has a higher tail index than $X$, then red returns will beat green returns with probability one for large $S$. 

\subsection*{Estimation}
To estimate the distributions for the ethical and unethical strategies, we suppose that the $k$ sampled strategies with the highest risk-adjusted returns have been divided into $k_R$ unethical and $k_G$ ethical strategies, with respective returns $r_1,\ldots, r_{k_R}$ and $g_1,\ldots, g_{k_G}$, and we denote by $u$ the largest sampled return that is not among these $k$.  In our asymptotic framework the generalized Pareto distribution (GPD)~\citep{Davison.Smith:1990} provides a suitable probability model for $r_j-u$ and $g_j-u$, i.e., the `excess' returns over $u$.  The probability density functions  for the red and green excesses are
$$
\dfrac1{\tau_R}\left(1+\xi \dfrac{r_j-u}{\tau_R}\right)_+^{-1/\xi-1}, 
\quad 
\dfrac1{\tau_G}\left(1+\xi \dfrac{g_i-u}{\tau_G}\right)_+^{-1/\xi-1}, 
$$
for $j=1,\ldots, k_R$ and $i=1,\ldots, k_G$. 
%negative log likelihood
%$$
%- k_R\log\tau_R -\left(1+\dfrac1\xi\right)\sum _{j=1}^{k_R} \log\left( %1 + \xi\dfrac{r_j-u}{\tau_R}\right)_+
%$$
The shape parameter $\xi$ is the same as in~\eqref{GEV.eq}, and $\tau_R,\tau_G>0$ are scale parameters.  The effect of changes in both $\Delta$ and $\gamma$ appears in the ratio $\tau_R/\tau_G$, which will be larger than unity if there is an advantage for red returns, whereas  $\xi$ should be the same for red and green subsets.  This last property is helpful: $\xi$ can be hard to estimate from small samples, but inference for it will be based on all $k$ of the largest returns. 
The adequacy of the GPD is readily checked using standard techniques \citep[Ch.~4]{Coles:2001}, and the parameters can be estimated, and models compared, using standard likelihood methods \citep[Ch.~4]{Davison:2003}.

Having obtained estimates $\hat\xi$, $\hat\tau_R$ and $\hat\tau_G$, we estimate $p_U$ by Monte Carlo simulation as follows.  We generate standard uniform variables  $U_1^*,\ldots  U_R^*$ and Poisson variables $N^*_1,\ldots, N^*_R$ with mean $r_{k_R}$, all mutually independent.  We then compute $M_r^*=\hat\tau_R[\{1-(U_r^*)^{1/N_r^*}\}^{-\hat\xi} -1]/\hat\xi$, for $r=1,\ldots, R$, and estimate $p_U$ by
$$
\hat p_U = R^{-1}\sum_{r=1}^R \exp[-r_g\{1-\hat F_G(M_r^*)\}],
$$
where $\hat F_G$ denotes the fitted cumulative distribution function for the green exceedances over $u$, which is generalized Pareto with parameters $\hat\xi$ and $\hat\tau_G$. In the simulations described below we took $R=10^5$, which reduces variation in $\hat p_U$ to the third decimal place.

We performed a small simulation experiment to check these ideas.  For different settings with normal and $t_{12}$ returns, we simulated 10,000 samples, each with $S=10^4$ and $\eta=0.1$.  We constructed  each sample by generating $Z_1,\ldots, Z_S\iid F$, and then made red returns $\Delta+(1+\gamma)Z_1,\ldots,\Delta+(1+\gamma)Z_{S\eta}$, with the green returns being $Z_{S\eta+1}\ldots, Z_S$.  We took the $k=200$ largest returns for each sample, ascertained whether they were red or green, and obtained $u$, $r_1-u, \ldots, r_{k_R}-u$ and $g_1-u, \ldots, g_{k_G}-u$.  We then fitted the GPD to the entire sample of $k$ excesses, and to the red and green excesses separately, using a common value of $\xi$; this enabled us to compute the likelihood ratio statistic for testing whether $\tau_R=\tau_G$, based on the $k$ largest returns; the proportion of times this is rejected is the statistical power for testing the hypothesis $\tau_R=\tau_G$ at a nominal 5\% significance level.  If the return distributions differ greatly, then this power should be high.  We also computed the empirical value of $p_U$, based on whether the largest return in each sample was red or green, which would not be useful in practice, as it would equal either 0 or 1, based on the single sample available.  As estimates of $p_U$ we computed the empirical proportion $p_U' = k_R/k$ and the estimate $\hat p_U$ described above, both of which would be available in practice.

\begin{table}%[tbhp]
\centering
\caption{Summary results from simulation study with $\eta=0.1$.  $p_U$, $p_U'$ and $\hat p_U$, shown as percentages, are respectively the probability that red beats green, the average estimate of $p_U$ based on the top $k$ values, and the average estimate based on fitting generalized Pareto distributions to the red and green values.  Power (\%) is the estimated power for detecting a difference between the red and green samples. See text for details. }
\label{tab.1}
\begin{tabular}{lll cccc}
&$\Delta$&$\gamma$ &$p_U$ &$p_U'$&$\hat p_U$&Power\\
\midrule
Normal&  0&  0& 10.2 &10.0& 13.4&  ~5.9\\
&  0.5 & 0 & 41.4 &25.7 &47.7& 19.3\\
&0  &0.2& 54.0 &20.0& 57.5& 46.4\\
& 0.5&  0.2 &86.8& 38.6& 90.3& 90.4\\
$t_{12}$&0&0&~9.8 &10.0&12.8&  ~5.2\\
& 0.5 & 0&  20.4 &21.3 &25.4&  ~5.4\\
&  0&  0.2 & 33.7& 18.3& 37.6& 20.1\\
& 0.5& 0.2 & 50.1 &32.1& 58.4& 33.0\\
%     [,1] [,2] [,3]  [,4]  [,5]  [,6]  [,7]
%[1,]  0.1  0.0  0.0 10.22  9.98 13.36  5.87
%[2,]  0.1  0.5  0.0 41.36 25.74 47.73 19.24
%[3,]  0.1  0.0  0.2 53.97 19.98 57.54 46.38
%[4,]  0.1  0.5  0.2 86.82 38.64 90.25 90.35
%[5,]  0.1  0.0  0.0  9.82 10.00 12.78  5.19
%%[6,]  0.1  0.5  0.0 20.35 21.28 25.24  5.41
%[7,]  0.1  0.0  0.2 33.66 18.26 37.56 20.07
%[8,]  0.1  0.5  0.2 50.10 32.10 58.41 32.97\bottomrule
\end{tabular}
\end{table}

Table~\ref{tab.1} summarises the results of this experiment.  The rows with $\Delta=\gamma=0$ show that $p_U$ and $p_U'$ are both close to the expected value of 10\% when there is no difference between red and green returns, and the power is close to the anticipated value, 5\%.  Although $p_U'$ increases when either of $\Delta$ or $\gamma$ is positive, it generally has a downward bias, and $\hat p_U$ appears to provide a better estimate of $p_U$.  On the other hand computations not shown indicate that $\hat p_U$ can be highly variable, though taking $k=500$ reduces its variance.  The power increases when $\Delta$ or $\gamma$ is positive, as predicted by the asymptotic theory; the power shows that when $\Delta=0.5$ and $\gamma=0.2$, for example, a difference between red and green returns can be detected in around 91\% of samples.  For the $t_{12}$ returns,  $p_U$ and its estimates again increase, but more modestly, and more for increased volatility, $\gamma>0$, than for increased mean, $\Delta>0$.  Again, this corresponds to the asymptotic theory.

%\b{In most cases the estimate of $\upsilon$ based on the $k$ largest returns is substantially lower than the value based on the maximum return, and the corresponding estimate of $p_U$ may also be a large underestimate of the true value.  I fear this knocks a hole in the proposed estimation strategy in the main paper.}

%\subsection*{Other stuff}
%\textbf{Asymptotic approximations for special cases, and an explanation of the growth of $\upsilon$ as a function of $S$ in the Gaussian case?}

\subsection*{Computation of $p_U$}
Let $m=S\eta$ and $n=S(1-\eta)$.  It is straightforward to check that 
$$
p_U = m\int F^n\{\Delta + (1+\gamma)x\}f(x) F^{m-1}(x)\,\D{x}, 
$$
which can be estimated by Monte Carlo simulation as follows:  
\bi
\item generate $U_1,\ldots, U_R\iid U(0,1)$, then set $M^*_{r} = F^{-1}(U_r^{1/m})$ for $r=1,\ldots, R$;
\item compute an  estimate 
$$
p_1^* = R^{-1}\sum_{r=1}^R F\{\Delta + (1+\gamma)M_{r}^*\}^n
$$
of $p_U=\pr(M_G\leq M_R)$; 
\item repeat the steps above, with $U^*_r$ replaced by $1-U_r^*$ to give an estimate $p_2^*$;
\item return $p_U^*=(p_1^*+p_2^*)/2$ as an estimate of $p_U$.
\ei
The first step uses inversion to generate maxima $M_{r}^*$ directly from $F^m$, the second step averages the exact probabilities $\pr(M_G<M^*_{r})$, and the third and fourth steps use antithetic sampling to reduce the variance of $p_U^*$. 
With $R=10^5$ this gives probabilities accurate to three decimal places almost instantaneously.  The R \citep{R2018} code below embodies this.
\begin{verbatim}
prob.sim <- function(S, eta, delta, gamma, R=10^5)
{ # F is distribution function and Finv its inverse
  n <- (1-eta)*S
  m <- eta*S
  u <- runif(R)
  x <- Finv( u^(1/m) ) 
  m1 <- mean( F(delta+(1+gamma)*x)^n )
  x <- Finv( (1-u)^(1/m) )
  m2 <- mean( F(delta+(1+gamma)*x)^n )
  (m1+m2)/2
}
\end{verbatim}
High-precision arithmetic may help in computing $p_U^*$ more accurately for very large $S$, though its precise value is rarely crucial.

\subsection*{Infinite strategy spaces and correlated returns}

As one example of the kind of approach discussed in the paper, consider the following:

Let $C(u,v)$ denote the copula that determines the dependence of random variables $U$ and $V$ having uniform marginal distributions.  One standard measure of extremal dependence is \citep{Coles.Heffernan.Tawn:1999} 
$$
\chi(u) = \pr(U>u\mid V>u) = \dfrac{1-2u + C(u,u)}{1-u}, \quad 0<u<1,
$$
where $u\approx 1$ is of most interest in the present context.  If $\chi = \lim_{u\to 1} \chi(u)>0$, then $U$ and $V$ are said to be asymptotically dependent, with $\chi=1$ corresponding to total dependence and $\chi=0$ to so-called asymptotic independence. The quantity $2-\chi$ can be roughly interpreted as the equivalent number of independent extremes at high levels of $(U,V)$, so $\chi=1$ yields one `equivalent independent' variable, and $\chi=0$ yields two `equivalent independent' variables.  Rank-based estimators for $\chi(u)$ from independent data pairs $(u_1,v_1),\ldots, (u_n,v_n)$ are available for high values of $u$, e.g., $u=0.95$.  As these are based on the ranks, the marginal distributions of $U$ and $V$ are irrelevant.

To apply these ideas, suppose that $A(s)$ can be treated as a stationary process, that there is a measure of distance on $\calS$, and evaluate $A(s)$ on an equi-spaced  grid, at $s\in 0\pm \delta, \pm2\delta, \ldots$, say.  Thus we can observe the joint properties of $A(s)$ at distances $\delta, 2\delta, 3\delta$ and so forth, taking $U=A(s)$ and $V=A(s+k\delta)$ for each $s$ in the grid.  If we take all such distinct pairs a distance $k\delta$ apart and estimate $\chi(0.95)$ as described above, then we can assess the dependence of the extremes of the process at lag $k$, for example by plotting  the estimate $\hat \chi_k$ against $k\delta$. This extremogram \citep{Davis.Mikosch:2009:Ext} will equal unity for $k=0$, and should drop to zero as $k$ increases, and thus can be used to assess the approximate number of equivalent independent values in $\calS$. 

\begin{figure}[t] %  figure placement: here, top, bottom, or page
   \centering
   \includegraphics[width=3in]{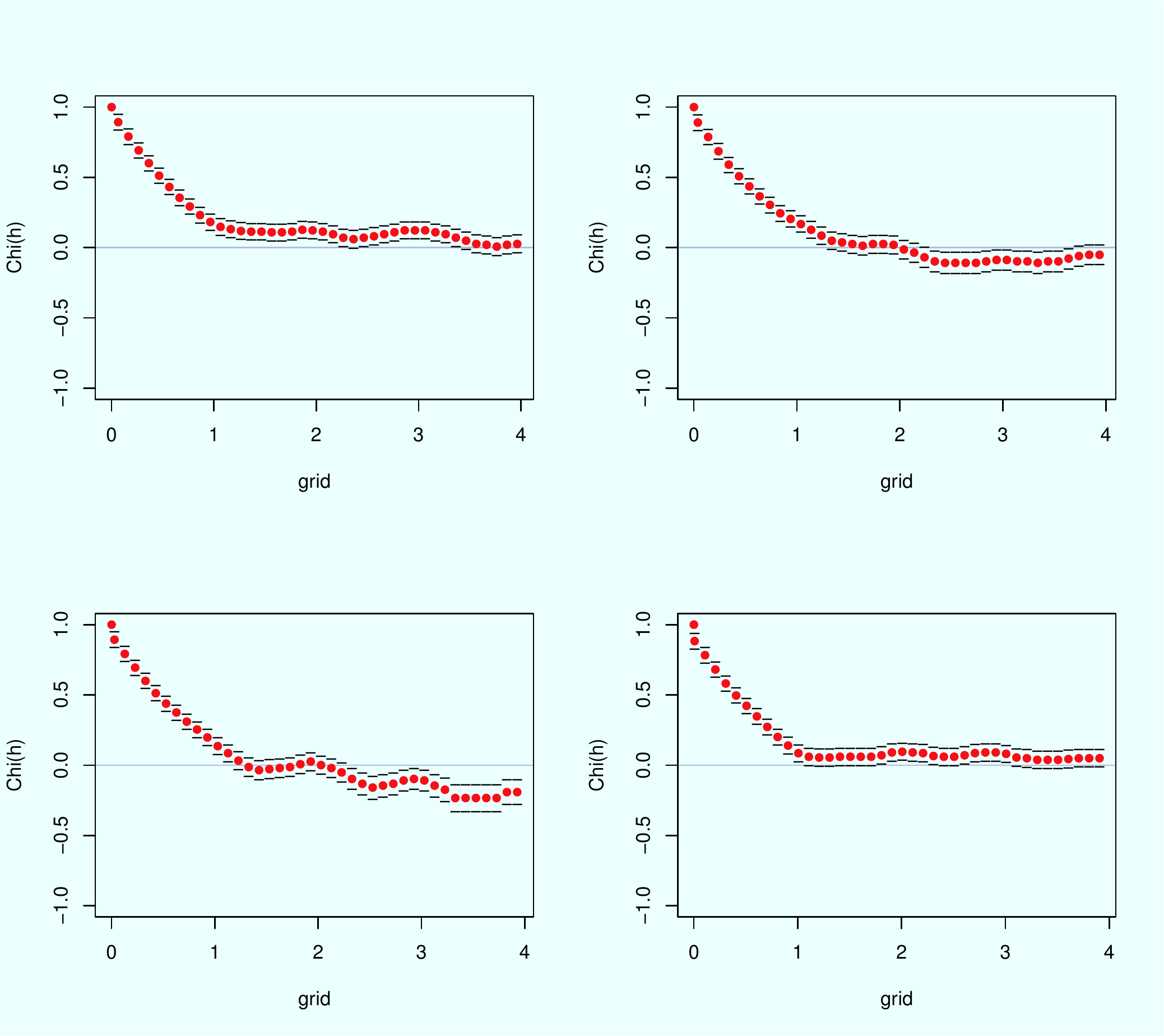} 
   \caption{Four examples of $\chi_k$ for the linear interpolation process described in the text.  The red points show the estimates of $\chi(0.95)$ at different lags, and the tick marks show 95\% confidence intervals for individual estimates.  The sharp initial decline shows that local dependence of extrema of $A(s)$ becomes negligible when $k\delta>1$ or so, as would be expected from the construction of $A(s)$. }
   \label{chiplots}
\end{figure}

To illustrate this, we took $\calS=[0,1000]$, created a function $A(s)$ by linear interpolation between $S=1001$ independent Gaussian variables at $s=0,1,\ldots, 1000$, and evaluated $A(s)$ on a grid with random initial value and $\delta=0.1$.  Figure~\ref{chiplots} shows these plots for four simulated functions.  The sampling properties of $\chi_k$ for $k$ large mimic those for the usual time series correlogram in the presence of strong dependence and are not good, but the sharp decline near the origin shows precisely the behaviour we expect; it appears that extreme values of $A(s)$ would be independent of those for $A(s\pm2)$ or perhaps $A(s\pm1)$, as we would anticipate from its construction. Thus if we sampled $\calS$ at sites no closer than two units apart, the corresponding values of $A(s)$ could be taken as independent at extreme levels. 

Although further refinement is certainly feasible, the discussion above strongly suggests that it should be possible to identify an approximate number of `independent' extrema in an infinite strategy space, under assumptions similar to those above, perhaps using a development of the ideas in Leadbetter~\cite{Leadbetter:1991}.

\end{document}